\documentclass[aps,twocolumn]{revtex4}

\usepackage{epsfig}
\usepackage{graphicx}
\usepackage{color}
\usepackage{subfig}

\begin{document}
\renewcommand{\thefigure}{\arabic{figure}}
\setcounter{figure}{0}

\bibliographystyle{apsrev}

\title{Fast Fisher Matrices and Lazy Likelihoods}

\author{ \surname{Neil} J. Cornish}
\affiliation{Department of Physics, Montana State University, Bozeman,
MT 59717}

\begin{abstract}
Theoretical studies in gravitational wave astronomy often require the calculation
of Fisher Information Matrices and Likelihood functions, which in a direct approach
entail the costly step of computing gravitational waveforms. Here I describe an
alternative technique that sidesteps the need to compute full waveforms, resulting
in significant computational savings. This approach can be used to greatly speed
up Bayesian inference applied to real gravitational wave data.
\end{abstract}

\maketitle


Over the past two decades there have been literally hundreds of papers written
describing parameter estimation studies in gravitational wave astronomy.
(we have to something to pass the time while waiting for the first detection).
See Refs.~\cite{Echeverria:1989hg,Finn:1992wt,Cutler:1994ys,Cutler:1997ta,
Flanagan:1997kp,Christensen:1998gf} for some important early examples.
The procedure is as follows: a waveform family $h_+,h_\times$ describing the radiation
produced by some potential source of gravitational waves is chosen,
and the detector response function is specified. From these inputs the waveform templates
$h({\bf x})$ describing the signals produced in the detector for model
parameters ${\bf x}$ can be computed. The output of the detector - or a network of
detectors, it makes no difference - can then be written as $s(t) = h({\bf x}, t) + n(t)$,
where $n(t)$ is the instrument noise. For theoretical studies it is usually assumed
that the noise is Gaussian with a colored spectrum $S_n(f)$, making it advantageous to shift
the analysis to the Fourier domain where the noise samples are uncorrelated.
The question is then asked, how well can the parameters ${\bf x}$ be constrained by
the data? The answer follows from considering how well waveforms $h({\bf y})$ with
parameters ${\bf y}$ are able to fit the data. The goodness of fit is found by taking
the squared difference between the model and data, scaled by the noise level:
\begin{equation}
\chi^2({\bf y}) = (s-h({\bf y})\vert s-h({\bf y})) \, ,
\end{equation}
where the brackets denote the standard noise weighted inner product
\begin{equation}
(a\vert b) = 2 \int \frac{a(f) b^*(f) + a^*(f) b(f)}{S_n(f)} \, df \, .
\end{equation}
The likelihood that the data would arise from a signal with parameters ${\bf y}$ in Gaussian
noise is then~\cite{Finn:1992wt}
\begin{equation}
p(s\vert {\bf y}) = C e^{-\chi^2({\bf y})/2} \, ,
\end{equation}
where $C$ is a constant that does not depend on the signal. The posterior probability,
$p({\bf y}\vert s)$, is simply the product of the prior probability, $p({\bf y})$,
and the likelihood, $p(s\vert {\bf y})$, divided by an overall normalization
factor (the evidence). If the noise is non-Gaussian, alternative expressions for
the likelihood function need to be used~\cite{Littenberg:2010gf}.

The parameter recovery accuracy is computed by looking at contours of the
posterior. For strong signals the central limit theorem says that the posterior distribution
is well approximated by a multivariate Gaussian distribution:
\begin{equation}\label{multi}
p({\bf y} \vert s) \simeq \frac{1}{{\rm det}(\Gamma/2\pi)}
 \exp\left(-\Gamma_{ij}\Delta x^i \Delta x^j/2\right) \, ,
\end{equation}
where $\Gamma_{ij}$ is the Fisher information matrix and $\Delta x^i = y^i - x^i$. It follows
that the Fisher matrix is given by the expectation value of the
negative Hessian of the log posterior density:
\begin{equation}
\Gamma_{ij} = -\langle \partial_i \partial_j \ln p({\bf x} \vert s) \rangle \, 
= (h_{, i}\vert h_{, j}) - \partial_i \partial_j \ln p({\bf x})\, .
\end{equation}
For sufficiently large signal-to-noise ratios, the variance-covariance matrix
$C^{ij}=\langle \Delta x^i \Delta x^j \rangle$ is given by the inverse of the
Fisher information matrix. Better error estimates can be found by directly estimating the
posterior distribution function using Markov Chain Monte
Carlo~\cite{Hastings:1970,Andrieu:2003} or Nested Sampling~\cite{skilling:2004,Feroz:2007kg}
techniques. Either way, these parameter estimation studies require that we
compute a large number of noise weighted inner products, $(a\vert b)$, which would
seem to imply that we need to compute a large number of waveforms. But that turns
out not to be the case.

Suppose that we want to map out the posterior using a Markov Chain or Nested Sampling.
For the moment I will ignore the noise term $n(t)$ to simplify the discussion, but later I will
show how to handle instrument noise. In theoretical studies it is often better
to leave out explicit noise realizations as there what we are interested in is the noise-averaged
performance~\cite{Nissanke:2009kt}, which is set by the noise level $S_n(f)$
in the noise weighted inner product. When $n(t)=0$ the quantity we need to compute is
$\chi^2({\bf y}) = (h-h' \vert h-h')$ where I'm using the shorthand $h=h({\bf x})$ and
$h'=h({\bf y})$. Suppose for the moment that we happen to have stationary phase approximation
waveforms for $h$ and $h'$ - I'll deal with time domain waveforms in a moment. Writing
$h = A(f) \exp(\Phi(f))$ and $h' = A'(f) \exp(\Phi'(f))$ we get
\begin{equation}
\chi^2({\bf y}) = 4 \int \frac{A^2 + {A'}^2 - 2 AA'\cos(\Delta\Phi)}{S_n}\,
df \, ,
\end{equation}
where it is understood that all of the quantities are frequency dependent.
The first two terms in the integrand are always slowly varying function of frequency,
and these terms can be integrated using a small number of function calls. The
oscillatory term that comes from
$(h\vert h')$ deserves more attention. So long as we are near maximum likelihood, the
phase difference $\Delta \Phi = \Phi(f)-\Phi'(f)$ will be a slowly vary function of
$f$, as will $\cos(\Delta\Phi)$, and once again the integral can be computed to
the desired accuracy with very few function calls. As we move away from maximum likelihood
the phase difference grows and evaluating the integral takes more function calls. But we are
not very interested in regions with low likelihood. It is simple to show that the
variance of $\chi^2$ is equal to the dimension $D$ of the parameter space, and it follows
that a Markov chain will rarely accept moves to places with $\chi^2 > 3 D$. Even there
the phase evolution is not large,
and the likelihood calculation remains inexpensive. Of course, moves will be proposed
to locations with low likelihoods, and these locations do lead
to very rapid oscillations in the $(h\vert h')$ terms. Which is why we simply set
$(h\vert h')=0$ whenever the phase change across the band exceeds some threshold (say
tens of radians). The results of a MCMC or Nested
Sampling analysis done in this way
is indistinguishable from what you get when computing the likelihood directly.

The same technique can be used to compute the Fisher matrix if we write 
\begin{equation}\label{fish1}
(h_{,i} \vert h_{,j}) \simeq \frac{ (\Delta h_i \vert \Delta h_j)}{4\epsilon^i \epsilon^j} \, ,
\end{equation}
with (no sum on $i$)
\begin{equation}
\Delta h_i = h({\bf x}+\epsilon^i \hat{e}_i) - h({\bf x}-\epsilon^i \hat{e}_i) \, .
\end{equation}
Expanded out, the numerical central differences in equation (\ref{fish1}) lead to $2D^2+D$
inner products that have to be evaluated. Since the $\epsilon^i$ are by definition small,
the integrands are all slowly varying and can be computed to the necessary accuracy with
a small number of function evaluations. The calculation is even simpler if we
work directly with the derivatives of the amplitude and phase:
\begin{equation}\label{fish}
(h_{,i} \vert h_{,j}) = 4 \int \frac{A_{,i}A_{,j}+A^2 \Phi_{,i}\Phi_{,j}}{S_n} \, df \, .
\end{equation}
Again, all the quantities in the integrand are slowly varying functions of frequency,
and the integrand can be computed at little cost. This method of evaluating Fisher matrices
has the added advantage that it is numerically far more stable than directly taking
derivatives of the waveforms~\cite{curt}.

We have seen that the likelihood and its derivatives can be computed very efficiently when
stationary phase approximation waveforms are available and instrument noise can be neglected.
But these conditions are very restrictive - the stationary phase approximation does not always
apply~\cite{antoine}, and it is impossible to ignore the instrument noise when dealing with real data.
Even when the stationary phase approximation can be used to compute the Fourier transform of the
waveforms~\cite{Droz:1999qx}, the presence of noise leads to $(h\vert n)$ cross terms
in the likelihood, which are expensive to compute since they involve rapidly varying integrands.

In what follows, I describe a new approach to computing the likelihood that can be used with time domain
and frequency domain waveforms in the presence of instrument noise. In essence the approach is based
on the heterodyne principle. That is, if we have signals $h({\bf x})$ and $h({\bf y})$ that differ by a small
phase difference, their product yields a low frequency beat signal plus a high frequency
signal that can be discarded without loss of information. The frequencies of the
signals do not have to be constant for heterodyning to work. Heterodyning has been used in the
search for gravitational wave signals from known radio pulsars~\cite{Dupuis:2005xv},
and to simulate LISA observations of white dwarf binaries~\cite{Cornish:2007if}.
What apparently has not been realized before is that heterodyning can be used to significantly
speed up MCMC and Nested Sampling explorations of the posterior for signals embedded in
noisey instrument data. Suppose that the primary or a secondary mode of the
posterior ${\bf x}$ has been located by some search algorithm and you would now like to fully
map out the posterior distribution. Rather than work with the full signal $s(t)$, first
Fourier transform the data, whiten using the noise spectral density $S_n(f)$, and heterodyne
using the carrier phase $\Phi({\bf x}, f)$. The high frequency components of the data can
now be thrown away, and the noise weighted inner products can be computed using templates that
are heterodyned against the carrier phase. The bandwidth of the heterodyned signal that
needs to be kept depends on the details of the analysis, but the data volume will typically
be reduced by many orders of magnitude. To gain the full benefit from this approach the
heterodyned templates must be computed directly at low cadence using the phase difference
$\Delta\Phi(f)$ (or the equivalent in the time domain). Note that carrier phase that
beats with the signal at the primary maximum will also beat with the signal at the secondary
maxima. The likelihoods computed far from the maxima will not agree with those computed using
the full signal, but if the heterodyned signal is given sufficient bandwidth, the
differences will be small in regions with noticable posterior weight. Heterodying works
equally well with non-Gaussian noise, and all of the techniques I have described can be
generalized to speed up the computations of non-Gaussian likelihood functions.

While a direct numerical implementation of the heterodyne technique is possible, I prefer to use a semi-analytical
approach that is similar to the method used to simulate the LISA response to the gravitational waves emitted by
white dwarf binaries~\cite{Cornish:2007if}. I'll start with frequency domain waveforms then move on to time domain
waveforms - the derivation is almost identical in both cases. To avoid unnecessary complications in the notation I will limit
the discussion to signals with a single phase term. The calculation can easily be extend to signals with multiple harmonics.
In most instances the contribution from each harmonic can be computed separately as they have very small overlap.
\begin{equation}
h(f) = A(f) \exp(i \Phi(f)) \, .
\end{equation}
Consider the chi-squared we need to compute:
\begin{eqnarray}\label{chisx}
\chi^2 &=& (h+n - h' \vert h+n - h') \nonumber \\
&=& (h+n \vert h+n) +(h'\vert h')- 2 (h\vert h') - 2(h'\vert n) \,.
\end{eqnarray}
The first term, $(h+n \vert h+n)$, can be computed once and stored. It
never changes. The next two terms are what we considered earlier, and
are given by
\begin{equation}
(h'\vert h')- 2 (h\vert h') = 4 \int \frac{{A'}^2 - 2 A A' \cos \Delta\Phi(f)}{S_n(f)} \, .
\end{equation}
The key observation here is that all the terms in the integrand are slowly varying functions
of frequency, so the integral can be done very cheaply. The final term $-2(h'\vert n)$ involving the
noise has a rapidly varying integrand which is expensive to compute directly. However, we can
use the heterodyne idea to re-write the integral as the convolution of two terms: one that is expensive
to compute, but only has to be calculated once; and a second, slowly varying term, that has to be
calculated for each new waveform. With a
little re-arrangement we can write
\begin{equation}
h'(f) = \left[ \frac{A'(f)}{A(f)}e^{-i\Delta\Phi(f)} \right] h(f) = {\cal A}(f) h(f)
\end{equation}
where ${\cal A}(f)$ is a slowly varying function of frequency. Defining
\begin{equation}
\alpha(\tau) = \int {\cal A}(f)  e^{-i 2 \pi f \tau} df
\end{equation}
we have
\begin{equation}\label{main}
(h' \vert n) = \int \alpha(\tau) \beta(\tau) d\tau
\end{equation}
where
\begin{equation}
 \beta(\tau) = 2 \int \frac{ h(f) n^*(f) \exp(2\pi i f \tau) + {\rm c.c.}}{S_n(f)} df
\end{equation}
The expensive-to-compute $\beta(\tau)$ can be calculated once then stored.
The $\alpha(\tau)$ function has to be computed for each $h'$, but it is given by the Fourier transform
of a very slowly varying function, which can be computed cheaply using a very short FFT. Likewise,
the integral in (\ref{main}), which in practice becomes a sum, only involves a few dozen terms.

The time domain version is almost identical. Starting with the signal
\begin{equation}
h(t) = A(t) \cos\Phi(t) = \Re[\bar{h}(t)]\, ,
\end{equation}
where $\bar{h}(t) = A(t) e^{i \Phi(t)}$, we can write
\begin{equation}
h'(t) =   \Re\left[ \frac{A'(t)}{{A}(t)}e^{-i\delta \Phi(t)} \bar h(t) \right] \, .
\end{equation}
The original waveform and its Fourier transform have to be calculated as
per usual, but this is a one time cost. Let us write
\begin{equation}
\bar h(t) = \int {h}(f) e^{2\pi i f t} df
\end{equation}
and 
\begin{equation}
\Delta(\alpha) =\int    \frac{A(t)}{{A}(t)}e^{-i \delta \Phi(t)} e^{-2\pi i \alpha t} \, dt  \, .
\end{equation}
The Fourier transform of the (complexified) perturbed waveform is given by the convolution
\begin{equation}
\bar h'(f) = \int \bar{h}(f-\alpha) \Delta(\alpha) \, d\alpha \, .
\end{equation}
Terms in the likelihood such as $(n\vert h')$ or $(h \vert h')$ can then be
re-expressed as
\begin{equation}\label{time}
(a \vert h) = \int w(\alpha) \Delta(\alpha) d\alpha + c.c. \, 
\end{equation}
where
\begin{equation}
w(\alpha) = 2 \int \frac{a^*(f)\bar{h}(f-\alpha)}{S_n(f)}  \, df \, 
\end{equation}
One again, the expensive-to-compute $w(\alpha)$ can be calculated once and stored.
The $\Delta(\alpha)$ have to be computed for each new waveform $h'$, but this can
be done at very low cost since it is given by the Fourier transform of a slowly varying function.
The integral in (\ref{time}) is in practice a sum that only involves a few dozen terms.

The techniques that I have described dramatically reduce the cost of repeatedly computing the likelihood
in the neighborhood of the best-fit signal, and should allow for a wider adoption of Bayesian inference in
gravitational wave data analysis.

\section*{Acknowledgments}
This work was supported by NASA Grant NNX10AH15G and NSF Award 0855407.

\end{document}